# Book to the Future ¶

## a manifesto for book liberation

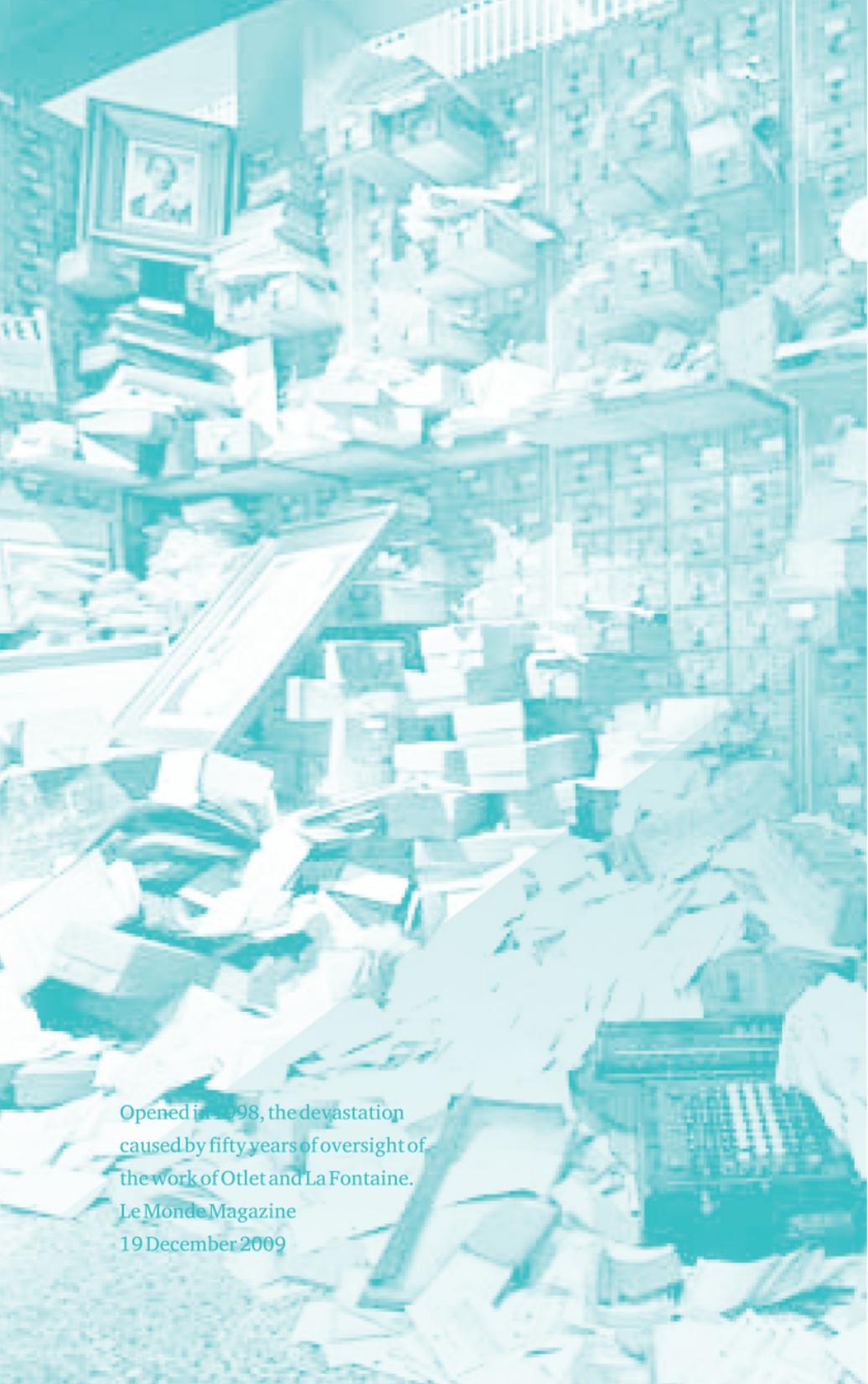

Opened in 1998, the devastation caused by fifty years of oversight of the work of Otlet and La Fontaine.
Le Monde Magazine
19 December 2009

# Book to the Future

## a manifesto for book liberation

Trial media referencing system in use. The HPC is using the Github implementation of Secure Hash Algorithm (SHA) version number, with the addition of a further media specific identifier. For this book publication it is the first edition print page number. For other media types the identifier could be a media specific identifier, for example a timecode for the case of audio or video publications. The HPC considers this trial citation identifier as an emergent *de facto* standard. The historical precedent being in the commentaries on Plato following the 1578 Stephanus references convention of citation.
See: http://plato-dialogues.org/faq/faq007.htm and in visual form http://plato-dialogues.org/stephanus.htm

SHA: 53ebf37188325b0787570d6bf74dc7dc9deec643
An example reference for this page is 53ebf37188 2

https://github.com/consortium/hybrid-publishing-research/blob/master/dist/docs/book_liberation_manifesto/Book_Liberation_Manifesto

**TABLE OF CONTENTS**

### 1. INTRODUCTION

The Hybrid Publishing Consortium (HPC) is a research network which is part of the Hybrid Publishing Lab and works to support Open Source software infrastructures. The HPC wishes to present practical solutions to the problems with the current stage of the evolution of the book. The HPC sees a glaring necessity for new types of publications, books which are enhanced with interfaces in order to take advantage of computation and digital networks. The initial sections of this manifesto will outline the current problems with the digital development of the book, with reference to stages in its historical evolution. We will then go on to present a framework for dealing with the problems in the later sections.

Now that there are floods of Open Access content for users to sort through, the book must develop to take on fresh interface design challenges – for improving reading, but also to support a wide range of communities. The latter include art, design, museums and the Digital Humanities groups, for all of whom video, audio, hyper-images, code, text, simulations and game sequences are needed.

HPC's view is that current technology provisions in publishing are costly, inefficient and need a step-up in R&D. To support technical, open source infrastructures for publishing we have identified the 'Platform Independent Document Type' as key. Our objective is to contribute to the working implementation of an open standards based and transmedia structured document for multi-format publishing. With structured documents and accompanying systems publishers can lower costs, increase revenues and support innovation.

HPC is about building public open source software infrastructures for publishing to support the free-flow of knowledge – aka book liberation. Our mission statement is:

> 'Every publication, in a universal format, available for free in real-time.'

This is our reworking of Amazon's mission statement for its

Kindle product:
> 'Every book ever printed, in any language,
> all available in less than 60 seconds.'

Currently digital publishing is dead in the water because for digital multi-format publications prohibitive amounts of time and costs are needed for rights clearance: the permissions required for each new format, the necessary signed contracts etc. So something has to give. For the scholarly community, Open Access academic publishing has fixed these problems with open licences, but other publishing sectors outside of academia remain frozen by restrictive licensing designed for print media.

Our efforts in building technical infrastructures will be wasted if content continues to be locked in, and this is where HPC's issue becomes as much a political as a technical problem. Open intellectual property licences, such as Creative Commons, are not enough on their own. Something else is needed if we want to support the free flow of knowledge: a way to financially support the publishers and the chain of skilled workers who are involved in publication productions. This can be either by a form of market metrics or by fair collections and redistribution methods, with the latter involving a little less fussing around than some market measurement. Open Access has meant publishers are still paid; it is simply that the point of payment has moved away from the reader to another point in the publishing process, where the free flow of knowledge is not hampered.

## 2. BROKEN WORKFLOWS

Publishing is the largest creative industry in terms of revenue. For example, in the EU there are 64,000 publishers with total annual revenues of 23€ billion.[1] The top 20% of publishers generate 80% of the revenues which, if EU figures are taken as a guide, means they are slicing off a cool 18€ billion annually. In the EU wealthier publishers can afford digital workflow systems which are prohibitively expensive for others, starting at 100,000 euro per annum in end-to-end costs. The 51,000 publishers who make up bottom 80%, with average revenues of less than three million euro, get by with various hand cranked custom solutions. It is these smaller publishers and all the self publishers, authors and institutions that we need to help.

A high end digital publishing system involves workflow integration and dynamic publishing features: multi-format publishing; standardised markup; rights management; asset management; reading metrics; automated distribution; metadata management; revisioning; document management; and payment systems, etc. It is notable that high end publishing systems continue to rely for many of these processes on offshored cheap labor.

Take one part of the workflow, multi-format digital publishing, which involves publishing to eBook, HTML, PDF, App and XML or other markup. Each of these formats has to be 'Publication Ready Output' for each distribution channel, which involves more than merely making the appropriate file type per format. We can see that the problems here are multi-format design layouts and revisioning. Currently a typical publisher would use a tool chain most likely comprising Microsoft Word and Adobe Creative Suite, neither of which are capable of making layout designs and handling revisions for multi-format in any practical or efficient manner. In this conventional scenario the workflow for each format requires a separate workflow for layout design, adding a new cost overhead for each format. And then on the side of revisioning, for example adding last minute edits, the current tool chain, again, involves each format being a separate workflow, so that updating a simple typo means editing four or five different files, which adds costs and drives editors crazy. These two factors alone – out of many more – make the digital publishing workflow uneconomic and unviable for the publisher. The net result is that publishers miss out on revenues and find it nearly impossible to entertain thoughts of innovating their processes or product lines.

There are many new online services with better and more integrated workflows, but they need more support in terms of development to reach maturity before publishers switch systems. The risks to the publishers of these new services is they will close down, due to insufficiently robust technology, or because of other problems, which means they are not viable for an industry with hard and fixed deadlines.

It is not solely key applications that are letting publishers down, it is also standards and technologies. It remains the case that common standards for document markups like HTML and EPUB cannot properly cope with basic publication components, including footnotes, styling in running headers, pagination and annotation. Infrastructural technologies are also unavailable as public services for reuse by individuals or businesses: examples include public web search engine indexes, cost free micropayment and Optical Character Recognition. Each of these areas does feature attempts and programmes to address the issues, but these have serious flaws or unresolved problems. Specifically, a public search engine index is only just being proposed in the EU as the Open Web Index[2] as proposed by Dirk Lewandowski of the Hamburg University of Applied Sciences, for micropayments BitCoin remains unviable while its value is so volatile, due to lack of regulation over speculators, while OCR projects like Google's 'Tesseract OCR'[3] involve Google maintaining privacy over its word pattern recognition for scanning in Google Books.

Nevertheless there are many groups working on improvement on a variety of areas in the technology tool chain as public infrastructure: W3C, International Digital Publishing Forum (IDPF), research councils and knowledge infrastructure groups; Deutsche Forschungsgemeinschaft (German Research Foundation, or DFG), Jisc, foundations and, most importantly, Open Source initiatives (e.g., The Libre Graphics Meeting) and the startup sector.

A specimen sheet of typefaces and languages,
by William Caslon I, letter founder, c. 1728.
https://en.wikipedia.org/wiki/File:Caslon-
schriftmusterblatt.jpeg

### 3. THE POOR BOOK

Industry pressures have led digital publishing to create a poor simulacrum of the book form – notably the eBook, which degrades or completely loses the typographic or mnemonic qualities of the paper book: page number, folios, speed of browsing, typographic detail of fonts and kerning etc. The typographic, navigational and other conventions of moveable type print have been contributed over the centuries by many anonymous printers, clerics and publishers. The book has never been a fixed entity but instead has evolved, normally acquiring improvements in the process, yet in the technology environment of the last forty years this process seems to have been reversed. Looking at the typographic craft and art in the sample illustration of multilingual typesetting below – English, Hebrew, Greek and Arabic – which dates from 1728 and is by the letter founder William Caslon, it is clear that a current e-ink reader would be hard pushed to equal this level of typographic quality or, more specifically, to render the language glyph sets and the letter spacing to aid reading. So, once more, a company like Amazon might have a mission statement for its Kindle product, 'Every book ever printed, in any language, all available in less than 60 seconds', but the key questions are those of what they will look like and how they will work for the reader.

If our standpoint was that of four decades ago then the technology companies and research funders could be excused for not addressing the raft of outstanding fundamental technology-design issues concerning publishing, books and reading. Unfortunately these issues have been poorly addressed since then, despite the ensuing technological advances. In 1974 the basics of the personal computer, tablet and networking were still challenges only just being overcome in terms of processing power, technologies for high quality displays, functional programming languages, standardised protocols etc. But these issues were mostly resolved twenty years ago – and with Moore's

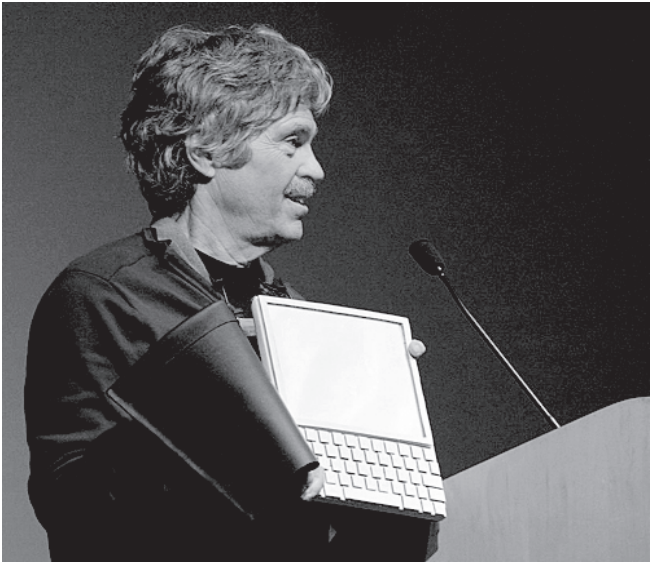

The DynaBook was first described by Kay in 1968 and then written up in a paper 1972, 'A Personal Computer for Children of All Ages'.

law of processor exponential improvement the future should not have been difficult to plan for.

Moving on from the basics of the book and looking at what the book could aspire to, what happened more than forty years ago alongside the invention of the personal computer by Alan Kay and the Learning Research Group (LRG) at Xerox Palo Alto Research Center (PARC), was the creation of the idea of the DynaBook (which first appeared in a paper entitled A Personal Computer for Children of All Ages).[4] DynaBook is short for 'Dynamic Book'. Essentially Kay and the team at LRG invented the personal computer and tablet, which Steve Jobs then copied and sold as the Apple Computer and, later, iPad. What failed to happen, and what Apple and others didn't pick up on, was the vision which lay behind Dynabook to enhance the book by understanding how people learn.

The DynaBook encapsulates an experience which is more active than passive, providing us with a better 'book'. Moreover for Kay the personal computer and ideas about what a book could become were rooted in an understanding of the technology based around McLuhanesque notions. Kay could see that industry trends led to computers being designed with much of their content adopted from previous media, the metaphor of the page in GUIs for example from print, with networked computers' own attributes only just beginning to be discovered.

Kay's other contribution to the idea of the book was the 'Active Essay', a publication that includes computational objects, that is, essays containing text and executable code to run simulations. One example is outlined in 'Active Essays on the Web', [Takashi Yamamiya, 2009
(http://www.vpri.org/pdf/tr2009002_active_essays.pdf);
Yamamiya has been a member of the Viewpoint Research Institute founded by Alan Kay.] The current web version of the Active Essay runs as Chalkboard
(http://tinlizzie.org/chalkboard/#Home).

Interestingly the technologies of the last eight years of a post-LAMP (Linux, Apache, MySQL and Perl) model – delivering Javascript and real-time browser updating and template libraries, like Angular.js, as well as template libraries like Bootstrap, NoSQL speedier scalable database delivery, together with instant cloud deployment – have meant that the ideas of the DynaBook and the Active Essay can start to come into play. This has been accompanied with a change in people's expectations towards demanding dynamic interface environments.

### 4. THE UNBOUND BOOK

Transmedia publishing in a post-Open Access context will include new kinds of media and masses of open licensed content (assuming the blockage of rights clearance is removed), with granular content reuse made possible from Open Access repositories. This has led us to re-examine the conventional book. To understand the book and translate it for computational use, we have found that we have had to atomise it, breaking it down into its smallest parts, before rebuilding a computable representation. This means creating a structured tree of components with meta descriptions, that is able to integrate many external and constantly changing data sources.

Once the book is broken apart then the publishing architecture that stems from the conventions of knowledge institutions and the divisions of labor come under examination. They include archives, education, research and library. With a free hand to recombine models from these areas of knowledge management, we see that the visions of the book developed in the past, as well as material from information science histories, lend inspiration when examining the basic limitations of current Internet technologies, HTTP and HTML etc. currently being adopted for the development of the digital book.

The development of the conventional book has been closely accompanied by parallel experiments with the unbound book form, essentially what became the library card catalog. In the 2011 book 'Paper Machines'[5] the media historian Markus Krajewski traces a history of the European unbound book beginning as library records in the sixteenth century, as created by Swiss librarian Konrad Gessner, on to Leibniz in the seventeenth century using the scholars' cabinet of quotes and references, the on to the US Dewey Decimal System of the nineteenth century and thence to transfer of the library record keeping system to businesses as the card index system in the early twentieth century. Key figures that bridge the transition from the 'universal paper machine' (Krajewski's term) to the 'digital

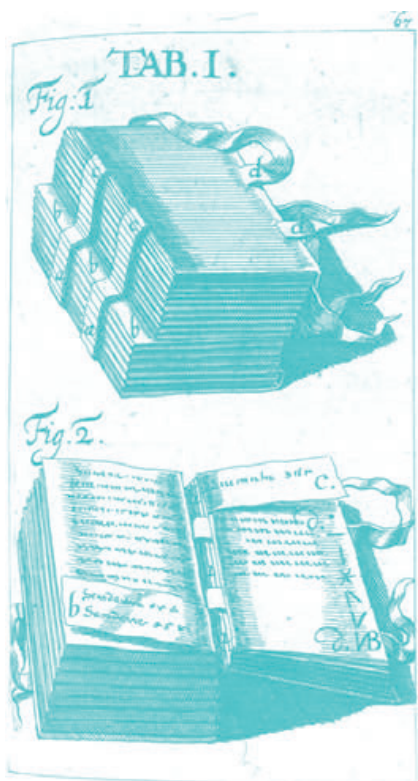

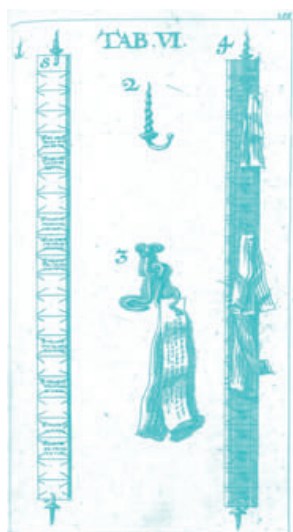

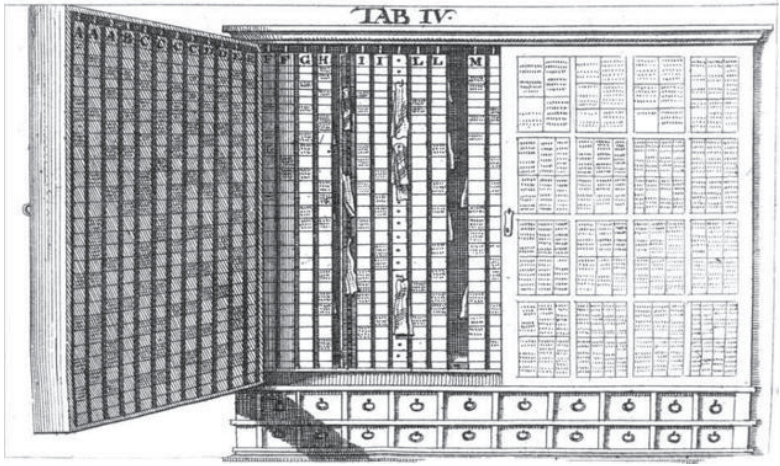

Top left: Hybrid card index in book form.
(From Placcius 1689, p. 67.)
Bottom left: The hook in the excerpt cabinet.
(From Placcius 1689, p. 155.)
Right: Leibniz 17C scholar's cabinet of quotes and references. Excerpt cabinet. (From Placcius 1689, p. 152.) Leibniz's method of the scholar's box combines a classification system with a permanent storage facility, the cabinet. So in a way this is similar to the use of Zotero or other citation management systems, but instead uses loose sheets of paper on hooks. The strips are hung on poles or placed into hybrid books.

universal machine' of the computer are Melvil Dewey and Paul Otlet. Dewey is renowned for his nineteenth century American dreams of universal access to knowledge. Less well known is the work of the Belgium librarian Paul Otlet in the early twentieth century with his pre-Internet, global paper packet Internet – or 'electric telescopes' as he described them – using telegraphs, early TV and audio radio.

This history helps in part to define the unbound book. To gain a fuller definition we need to add the agent of digital-disruption or – according to the term coined by economist Joseph Schumpeter – of 'creative destruction'.[6] 'Creative destruction' is a process in which a new economy emerges out of the destruction of a previous order. Technology innovation is the agent of this change, and Schumpeter describes the entrepreneur as the one who exploits it.

Ironically it is the unbound book and its prodigy, the card index system, that led to the punch card, the early data packet of what are now packet networks. The packet network is where all media can be broken down into common data packets and sent to any device. It is this technology that acts as the agent of creative destruction, that makes up basic Internet and mobile networks, and has made the concept of the unbound book finally realisable. It is the scaling up of this function of packet networks which means that the fundamentals of publishing are in flux. The innovation of the unbound book and the replacement of print books in many contexts means that economic models crumble, institutions of knowledge lose their relevance, and copyright laws become unworkable and act as an impediment to knowledge dissemination.

Dewey and Otlet both point to inspirational visions of the unbound book as the knowledge components of universal libraries, visions which embody ambitions to make the world's knowledge universally available. Dewey is more famous, with his mechanism of the classification system and the card system, immersed in the Taylorist obsessions of efficiency, speed and

THE UNBOUND BOOK

Top: Dewey's scheme, displayed by The Bridge.
(From Bührer and Saager 1912, p. 4.)
Bottom: Dewey's scheme, displayed by the Institut
International de Bibliographie. (From Institut
International de Bibliographie 1914, p. 45.)

BOOK OF THE FUTURE

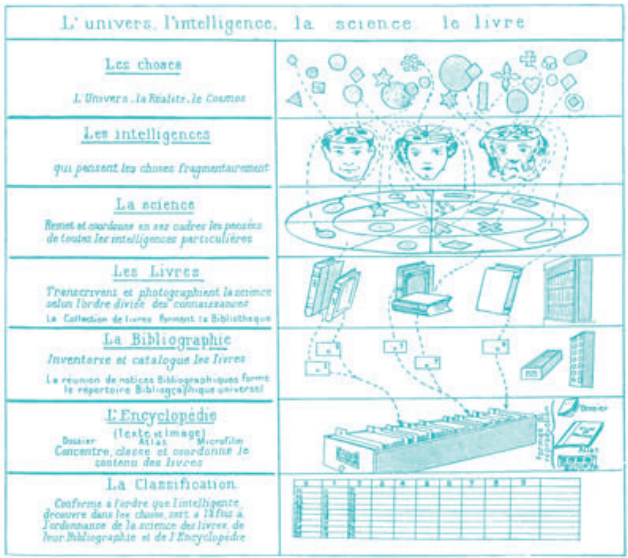

Top: Paul Otlet, Traité de Documentation, 1934, p.41.
Bottom: Paul Otlet, 1934, vision of universal knowledge systems.
http://mundaneumpaulotlet.tumblr.com/
http://www.flickriver.com/photos/marcwathieu/sets/72157623466540563/

economies of scale. Otlet was lost to obscurity in the chaos of World War Two, but had imagined and built elaborate index card knowledge system, again as a vision of the global library. Otlet has only recently been re-discovered, for example in the book by Alex Wright 'Cataloging the World: Paul Otlet and the Birth of the Information Age', 2014. [7]

## 5. DESIGNING THE BOOK OF THE FUTURE

If we think about new screen interfaces for publications, we can start by considering the readers and how to help them read: to assist them to remember, explore, experience, gloss, browse, reuse and rewrite. We can enhance what they may have already learned to do with paper books. The new interface, supported by real-time available open IPR content, can have references inline, with full copies of any publication mentioned, as well as highlighting of the sections the reader is interested in. In fact, any media cited or referenced should be available in full for transmedia publication, be it a game sequence, an exact time stamped point in a video clip, a scalable 3D model, a data calculation or a simulation, etc.

To recap, the most high profile impediment to the transmedia publication has been copyright. In the Open Access model there is no need to clear rights, because it has already been done via an open licence. Until this happens more generally transmedia publications will remain a non-starter. As well as abolishing rights clearance the point of payment must move up the value chain.

The two other hurdles to the transmedia publication are, first, reader expectations and, second, technology. Both have turned around one hundred and eight degrees since the advent of the precursors in the journey of the digital publication, Hypertext and Multimedia, which appeared in the 1980s and 1990s respectively. Now users expect real-time updating interfaces and are disappointed when when they are absent. Technologies for interfaces now have Javascript for rich interactivity, design frameworks are templated, and standards allow for system media transfer and communication automatically via Application Programme Interfaces (APIs).

If we are thinking about the new publication design, the reader as receiver or consumer is only one role to consider. We must also examine many of the other roles in the lifecycle of a publication: the librarian, writer, designer, editor, tutor etc.

As an example in the area of the writer and editor, real-time collaborative text editors – GDocs, Fidus Writer, Etherpad, Ethertoff – change the skill set of the user, change the interface of the publication from read only to read/write, and so intervene in the intimacy of the act of authoring. As Kenneth Goldsmith explored in his book *Uncreative Writing*[8], in this publication lifecycle there is also the role of machinic writing and interventions to consider. In the example of real-time collaborative texts the key algorithm is called Operational Transformation,[9] essentially storing all possible edits, in case they need to be retrieved by one of the collaborators.

For our purposes of designing interfaces for new types of publications we group such semi-automated computational processes in the digital workflow under the umbrella term 'Dynamic Publishing'; they include: layout, multi-format conversion, distribution, rights management, file transfer, translation workflows, document updates, payments and reading metrics etc. Our aim is to explore these processes in rethinking the publication interface.

As part of Dynamic Publishing and the networked publication, privacy has to be addressed as a fundamental right of the reader. At the same time, tracking and the reading equivalent of the 'social graph' are qualities that are very useful. Nevertheless, privacy needs to be addressed technically and politically. Firstly, metrics data needed for a 'reading graph' must be anonymized. Secondly, access must be allowed to the missing matrix of uncreative publishing: the Big Data of reading analytics, incorporating the Ngram of reading patterns.

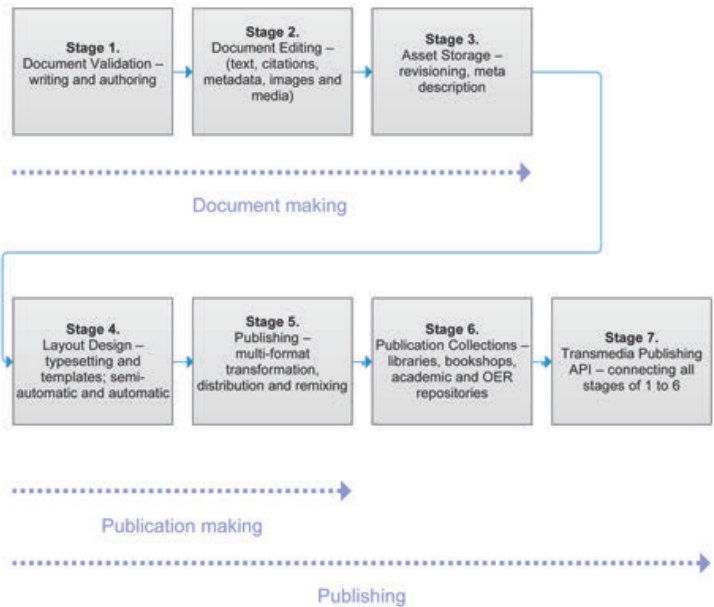

Major categories in the publishing infrastucture.

### 6. INFRASTRUCTURE

The objective of the Hybrid Publishing Consortium is to support Open Source public infrastructures for transmedia, multi-format, publishing. This means using structured document architectures to output publication formats such as EPUB, HTML5, ODT, DOCX, screen PDF and PDF for print-on-demand etc.

All of this can be achieved by connecting existing platforms and supporting development communities with expertise, resources, a knowledge network and by building new components if they are missing.

Our approach is format agnostic, so platforms can use XML, HTML, Markdown, ODT etc. for document markup because we would support an API for interoperability, so long as the formats support the required features for what we call 'Publication Ready Outputs' (PROs).

A PRO is made up of a combination of file type specifications, metadata requirements for the distribution channel, as well as 'style guides' for editors and designers for creating specific publication components for multi-format publishing. The latter includes items such as tables of contents, and front cover and back cover texts. A PRO profile is needed for each output format because one format will not automatically translate to another format, e.g. a print book to EPUB. Fundamental to this is definition of a Single Source file, which will act as a master universal document and also a container for multiple sources of external data, for example different image sizes from an external source for responsive web design, or external metadata, such as bibliographic citations or book trade metadata for publishing purposes.

However this technology stack is now being superseded by javascript technologies creating virtual machines, including items such as Node.js, unstructured databases like MongoDB[10] and real-time display technologies like Meteor. The result is a smoother GUI experience in which content from multiple source is updated in real-time in the browser. This is a move away from the server-based, client architecture of LAMP. For the Open Access community this is an exciting opportunity because it means the Open IPR content repositories can offer a rich resource for all types of new publishing models, as well as Virtual Research Environments (VRE) for scholars. Open Source examples that have used these technologies can be seen in investigative journalism such as DocumentCloud.
(http://www.documentcloud.org)

### Major Stages in the Workflow

We have identified six stages in the publishing workflow/lifecycle.

#### I) Document Validation – writing, authoring and structuring

Validation is required to create the structured documents. An interactive feedback GUI is needed to gain the authors' help to make structuring decisions that the validation algorithm cannot take on its own. These involve the validation rule set, structure and semantic information: Document layout e.g. headers, bold etc; Document structure e.g. pagination, chapter etc; and Metadata fields and standards for the document. An external document editing system will be able to have our rule set applied to its documents, via an API.

#### II) Document Editing – text, citations, metadata, images and media

Adding more components to the document on top of the text document's linear string of text means that we need to be able to separate out these components, create a scheme for their storage, and allow access to external data and media sources. External sources would be citations from Zotero, as well as metadata from library system and archive repositories such as Pandora. Additionally, revisioning issues are important here.

### III) Asset Storage – revisioning, meta description

The publication asset management component of the technology stack will use a NoSQL based DB infrastructure, with metadata frameworks MODS and VRA Core4 – as used by the Tamboti metadescription framework of the Heidelberg Research Architecture.

### IV) Layout Design - typesetting and templates: semi-automatic and automatic

This involves the use of Multi-format Templates, which in turn requires connection to Content Distribution Networks (CDN), so that designers can author templates in software and graphic design libraries they are familiar with, like Bootstrap. In effect this means creating modified bootstrap modules for apps, mobile, EPUB etc. Examples would be open source frameworks such as PugPig (http://pugpig.com) and Famo.us (http://famo.us).

### V) Publishing – multi-format transformation, distribution and remixing

Multi-format transformation means using our own A-machine software eco-system for multi-format transformation. The end publications can then be distributed to POD and digital distributors and repositories via a number of aggregators. The structured document format will make the documents and publication available for remixing at a granular level, down to specific points in the text or video clips. The format is designed to allow a wide variety of new publication uses.

### VI) Publication Collections – library, bookshop, academic and OER repositories

Firstly, this means supporting the creation of collections of publications. It will include an API for distributing collection with Open Publication Distribution System (OPDS)[11] metadata for inclusion in other systems. Secondly, it means creating easy to deploy real-time web platforms using Meteor and Node.js etc. to allow publishers to set up their own libraries, repositories and shops. With these two sets of framework options publishers, editors, educators and librarians can create custom packages to fit into existing systems or deploy web platforms if needed.

### VII) Transmedia Publishing API

An Application Program Interface (API) is the way in which our systems' modular components can communicate with other systems on the internet securely. This means that the functionality we are researching and developing – including validation, publication asset structuring, templates, and collections – can be integrated into the other systems we are connecting to.

### 7. THE PLAN

It is important to emphasize is that the HPC is not a fixed and finalised group and we are only at the beginning of forming the network. We want to invite more people to join. The plan is for long term collaboration with a network of stakeholders to support Open Source infrastructures for transmedia, multi-format, scholarly publishing. The objective is to put a reliable and trustworthy tool set in front of publishers, so that those publishers themselves can then start to innovate. This means a wholesale replacement of proprietary software applications, improvements in Open Source tools, open standards and formats, and the introduction of new interoperable systems where they do not yet exist – for example for micro-payments. We do acknowledge that the process towards software maturity is a long one, and that Open Source is merely a design and engineering methodology and not a guarantee of quality. LibreOffice is an example of such a tool in the infrastructure; it has taken more than fourteen years of work for LibreOffice to become a reliable replacement for Microsoft Word, but it is now stable software. LibreOffice is an example of Open Source reverse engineering, of figuring out how something works and building a clone. Other projects are more ground-breaking and have different sets of challenges, but again we see that they can become market leaders over time – for example the eBook manager Calibre.

    We recognise that this development and adoption of Open Source is a political issue which involves policy, economics and technology, and which needs multi-stakeholder agreement to move technology developments forward.

    Our plan is divided into three complementary areas of activity: research, open learning and ventures. These activities would be supported by the formation of two entities, firstly an Research and Technology Organisations (RTO), the 'Hybrid Publishing Consortium', with a series of academic institutions and other partners, second by private companies, currently including 'Infomesh Technologies'.

Research – current research focuses on issues involved in multi-format transformation: the creation of a structured document; layout design issues; and understanding the users and their skill sets in this area of the workflow. Research will continue in a number of ways: on dedicated software projects, either as collaborations or as networks with other academic partners, but also in industry contexts and on ventures. Our next areas of focus will be on interactive validation GUI for structured writing, as well as on real-time web GUIs and modular templated designs.

Open learning – to support the Open Source community we are developing a number of units dealing with Dynamic Publishing, for an open curriculum of Bachelors and Masters courses which is being discussed by members of Libre Graphics Meeting. This would also be implemented in consultation with The Open Syllabus Project (OSP) of Columbia University. The courses would be designed to work with programmes of the UN World Summit on the Information Society (WSIS), OER track.

Ventures – to support the long term sustainability of infrastructure components, projects need to move from research and into product focused development as well as support service provision. These ventures would be based on projects that are developed by the Hybrid Publishing Consortium or by partners. Additionally we would develop a series of regional business hubs for local service provision, and to act as knowledge networks for technologists and designers to pick up the tools we are supporting and run their own ventures. As an example we have joined the Open Invention Network (OIN),[12] which is supporting Linux developers and businesses by building a collective legal defensive solution against predatory and restrictive Patent protection.

### 8. OUR RESEARCH

We have combined a number of methods that fit under the umbrella of Design Research, all of which involve knowledge production through the process of making. These methods are: rapid prototyping; interviews; workflow and lifecycle mapping; discourse analysis, specifically TPINK (Technology, Power, Ideology, Normativity and Communication); collaborative authoring of manuals and guides; Somerville's software requirements process; publication forensics; publication prototype productions; Open Source software releases and open code review; and stakeholder consultations and knowledge network creation.

Rapid prototyping allows us to test out dynamic publishing opportunities and ways of integrating software into user workflows. Our findings demonstrate the need for a technical tool that lets publishers' workflows adapt easily to the demands of multi-format publishing. The rapid prototyping projects fall into two categories. The first is infrastructure software design in the area of multi-format, single source publishing transformation engines. Second is a series of publisher prototypes, which means working with publishers to make examples of digital publication productions.

Our research was initially outlined in a research plan in 2012. This runs until 2015 and will then be reviewed with new priorities in order to run for a further three years. The 2012 research plan can be found here:
Dynamic Publishing–New Platforms, New Readers!
http://www.consortium.io/research-plan

### Rapid Prototypes

### Infrastructure

TypeSetr – is an Open Source publishing software component for multi-format document conversion. It produces the following formats with automatic templated design layouts and format conversion: EPUB 3.0, HTML5, PDF and others.
Code - https://github.com/consortium/typesetr-converter
Demo - https://typesetr.consortium.io

A-machine – is a software ecology supplied by different providers to complete the major steps in the publishing workflow and connect our existing document structuring work. This includes: meta description frameworks; layout template designs; distribution and sales; Open Access and OER formatting; validation etc. http://a-machine.net

### Publisher Prototypes

• Merve Remix – in which we digitised Merve Verlag's back catalogue of 150 titles, and are making it available online for remixing. http://merve.consortium.io
http://www.consortium.io/merve-remix

• Museums and Post-digital Publishing – with Fotomuseum Winterthur, and the publication *Manifeste!* In which we examine how the high quality museum catalogue can be digitised and taken into open learning and other contexts.
http://www.consortium.io/fotomuseum

• Traces of McLuhan – a Media Sprint at the Marshall McLuhan Salon - McLuhan archive, where we create a transmedia trace of a user's journey through the archive, using Heidelberg Universities Tamboti platform and a second archive platform Pandora. http://www.consortium.io/traces-mcluhan

• Moos Verlag – where we look to engage a new community with a 1970s urbanism publishing collection. This involves book scanning and re-publishing titles free online.
http://consortium.io/moos-verlag

http://www.librarything.de/catalog/moosbeat&tag=Urban%2BDesign

### Publication Forensics

Here it is important to bear in mind our objective of making software infrastructures for publishing that are reliable, based on a free-flow of knowledge using Open Source methodologies, and cost effective for publishers. There are three components to this software design process: understanding the real world problem, making an imaginative leap and, finally, a precise weaving of the first two into the material of the process in order to reduce ambiguity through an iterative requirements building process. This is where 'Publication Forensics' has emerged as a practice in our research. So far it has guided several key projects. These have included immersion in hundreds of volumes of the Merve Verlag back catalog and manually reconstructing scanned texts back into a data object semantically resembling a book. Also notable has been compiling a lexicon of all scholarly publishing types known inside WikiPedia–Festschrift, Gloss, Leak, Liquid book, Ted Book etc.

https://en.wikipedia.org/wiki/User:Mrchristian%5CBooks%5CA_Publication_Taxonomy

### Research Publications

We have established a publishing programme for a number of reports, special dossiers, 'good practices' guides, manuals and reference materials, all of which can found on our GitHub repository as hybrid publications.

https://github.com/consortium/hybrid-publishing-research

1. A Publication Taxonomy, 2014.

https://github.com/consortium/publication-taxonomy

2. Book Scanning & Book Scanning Manual, 2014.

http://bookscanner.consortium.io

3. Structured document writing manual and style guide, 2015.

4. Workflows/lifecycles – see example 'periodic table' below, 2015.

5. Publication Ready Outputs – definitions and guides for multi-format publication output targets and design issues

6. Standards guide to structure documents for multi-format publishing, 2015.

7. Technical maps of publishing infrastructure software and systems, 2015.

Digital Publishing Periodic Table,
Simon Worthington, 2012.

### 9. CONCLUSION

The idea of the free circulation of knowledge guides the HPC and helps bind partners together in our collaborations. This is a suitable moment to reflect upon the Open Access (OA) movement in academic publishing and its progress in book liberation since the Budapest Open Access Initiative (BOAI) was launched in 2002.[13] OA has since been adopted as the norm in many jurisdictions, but vested interests are creating confusion and political difficulties. As recently as last year the Netherlands government took a hard line with Elsevier, by withdrawing all payment to the company unless it complied with the government's OA policies. Elsewhere, the UK has opted for sweeteners to the publishing industry under the Gold Open Access scheme advocated by the Finch Report,[14] under which researchers pay publishers for the right to publish through Open Access. Meanwhile, over in the US Elsevier pays lobbyists to tighten research copyright[15] (Lessig), which has lead Harvard Magazine to label academia 'The Wild West'[16] (Harvard Magazine) of publishing. What is important to keep in mind is not the staggering annual profits corporations make from publishing, although in the case of Reed Elsevier this is £826 million per annum (2013)[17] from academic publishing, specifically, Science Technology and Medicine (STM) - straight out of the public purse. The real issue is that human knowledge cannot be shared and used to benefit humankind, because it is important to remember that the result of payment of this near-1€ billion annually is that only a relative handful of people can read or use academic publications.

Moving on to looking at publishing in general to ask the question of how Open Access (AKA book liberation) can be mapped onto this varied and large industry is complex. In the EU alone publishing is the largest creative industry, with an annual turnover in the region of 23€ billion (2009).[18]

A functioning and equitable economy is needed to support 'free-at-the-point-of-reading' and 'free-to-re-use' publishing models. Corporate capitalism does nothing but skim off the profitable parts of the stack while imposing distribution monopolies, to leave the bulk of publishers to live with the constant drone of 'start-up', 'entrepreneurialism' and 'disruption', as their only strategies for finding some imagined and as yet unknown economic model - an ever elusive Eldorado. These capitalistic mantras, however thin they might wear, still keep the thinking on these issues confined.

    The Book Liberation Manifesto suggests two ways forward for publishing. Firstly, a redistribution of the profits by the top earners of the publishing industry to the lower rungs or for those top earners to pay for the release of publications into public circulation. Second, for Open Source publishing software to be treated as infrastructure and to receive the same funding that national broadcast networks receive, or for it to be maintained and enhanced in the way other kinds of basic infrastructure provision are supported. The result of such infrastructures providing low cost ways of reaching publics via digital channels would be that publishers could afford to experiment and innovate. Ironically Charles Babbage, the inventor of the first mechanical computer, identified these capitalistic traits and their limiting effect on publishing nearly two centuries ago in 1818, in a book chapter entitled 'On Combinations of Masters Against the Public'.[19] What Babbage showed, via detailed calculations of labour and materials, was that publishers were falsely inflating the price of books, putting them out of reach of the common people. To no one's surprise his book was banned by the publishing trade. Now that the descendants of Babbage's 'Difference Engine' are at our fingertips in the form of the modern computer it is time to take a lead from the computer scientist Alan Kay:

'The best way to predict the future is to invent it.'

Alan Kay (1971), at a meeting of PARC, Palo Alto Research Center.[20]

### REFERENCES

1. European Commission, Joint Research Centre. Institute for Prospective Technological Studies
http://is.jrc.ec.europa.eu/pages/ISG/documents/BookReportwithcovers.pdf
2. Open Web Index http://openwebindex.eu
3. Tesseract OCR https://code.google.com/p/tesseract-ocr/
4. A Personal Computer for Children of All Ages, Alan C. Kay, Xerox Palo Alto Research Center, 1972
http://www.mprove.de/diplom/gui/kay72.html
5. Paper Machines About Cards & Catalogs, 1548-1929. 2011 by Markus Krajewski, Bauhaus University, Weimar.
http://mitpress.mit.edu/books/paper-machines
6. Capitalism, Socialism and Democracy. 1942 by Joseph Schumpeter. http://digamo.free.fr/capisoc.pdf
7. Cataloging the World: Paul Otlet and the Birth of the Information Age', Alex Wright, 2014.
http://www.catalogingtheworld.com
8. "Book Details : Uncreative Writing," accessed February 4, 2015, http://cup.columbia.edu/book/uncreative-writing/9780231149907.
9. Operational Transformation algorithm
http://en.wikipedia.org/wiki/Operational_transformation
10. About MongoDB http://www.mongodb.org
11. Open Publication Distribution System (OPDS) http://opds-spec.org/specs/
12. Open Invention Network (OIN)
http://www.openinventionnetwork.com/
13. "Budapest Open Access Initiative" 2002
http://www.budapestopenaccessinitiative.org/read
14. Accessibility, sustainability, excellence: how to expand access to research publications (aka Finch Report)
http://www.researchinfonet.org/publish/finch/ 2012 UK OA policy report
15. Aaron's Laws - Law and Justice in a Digital Age. Lawrence

Lessig marked his appointment as Roy L. Furman Professor of Law and Leadership at Harvard Law School with a lecture titled "Aaron's Laws: Law and Justice in a Digital Age." The lecture honored the memory and work of Aaron Swartz, the programmer and activist who took his own life on Jan. 11, 2013 at the age of 26. A full transcript can be found here http://www.correntewire.com/transcript_lawrence_lessig_on_aarons_laws_law_and_justice_in_a_digital_age | https://www.youtube.com/watch?v=9HAw1i4gOU4
16. The "Wild West" of Academic Publishing http://harvardmagazine.com/2015/01/the-wild-west-of-academic-publishing
17. Reed Elsevier (2014). 2013 Annual Report. Retrieved March 18, 2014 from
http://www.reedelsevier.com/investorcentre/reports%202007/Documents/2013/reed_elsevier_ar_2013.pdf
18. European Commission, Joint Research Centre. Institute for Prospective Technological Studies
http://is.jrc.ec.europa.eu/pages/ISG/documents/BookReportwithcovers.pdf
19. On the Economy of Machinery and Manufactures. Charles Babbage (1832), Chapter XXIX - On Combinations of Masters Against the Public, pp. 258-270
https://archive.org/stream/oneconomyofmachi00babbrich#page/n19/mode/2up
20. "Don't worry about what anybody else is going to do… The best way to predict the future is to invent it. Really smart people with reasonable funding can do just about anything that doesn't violate too many of Newton's Laws!" — Alan Kay in 1971. Inventor of Smalltalk which was the inspiration and technical basis for the MacIntosh and subsequent windowing based systems (NextStep, Microsoft Windows 3.1/95/98/NT, X-Windows, Motif, etc…). http://www.smalltalk.org/alankay.html

### ACKNOWLEDGEMENTS

Thanks to my HPC colleagues, the Hybrid Publishing Lab, HPC partner organizations and to Peter Carty for copy editing.

Simon Worthington is a research associate at the Hybrid Publishing Lab which is part of the Leuphana University of Lüneburg Innovations-Inkubator.

### About the Hybrid Publishing Consortium

The Hybrid Publishing Consortium is a research group which supports Open Source software for publishing infrastructures.

The Hybrid Publishing Consortium model of software is built on user research and rapid prototyping. The architectural approach is modular, back end orientated, with an emphasis on application frameworks, ISO standards and interoperability between services and providers, as opposed to creating stand-alone web facing applications.

The Hybrid Publishing Consortium is a research group of the Hybrid Publishing Lab in collaboration with partners and associates. The Hybrid Publishing Lab is part of the Leuphana University of Lüneburg Innovations-Inkubator, financed by the European Regional Development Fund and co-funded by the German federal state of Lower Saxony. As an business incubator our research is conducted with industry partners and we are supported to create new startup business ventures. Currently HPC has one startup InfoMesh UG which is specialising in MLA (Museums, Libraries and Archives) publishing.

SOFTWARE STACK

**Software**
- A-machine
- Sublime Text 2
- Scribus
- Bootstrap
- Transpect
- Calibre
- Javascripts

**SaaS**
- Google Docs
- Github

**Standards formal and de jure (ISO, W3C, IETF, UTR etc)**
- HTML5
- CSS
- EPUB 2.1
- UPUB 3.0
- PDF
- Dublin Core
- BICS
- BISG
- ISBN
- XML
- SHA

**Standards de facto**
- ORCHID
- Print format - 'A' format
- 'pocket' size
- 178 x 111mm

**Fonts**
- Charis SIL - SIL (originally known as the Summer Institute of Linguistics, Inc.)
- Work Sans

**Distribution & platforms**
- Anagram via Mute Publishing
- Metamute.org
- A-machine - Research Viewer
- Lightningsource
- US/UK Printing
- Ingram Advance Catalog
- Nielsen UK Pub Web
- Amazon Pro Seller
- Amazon Kindle
- Ingram Spark
- Apple iBooks
- Google Play
- Issuu
- Github
- Aaaaarg
- Archive.org
- OpenLibrary
- LibraryThing

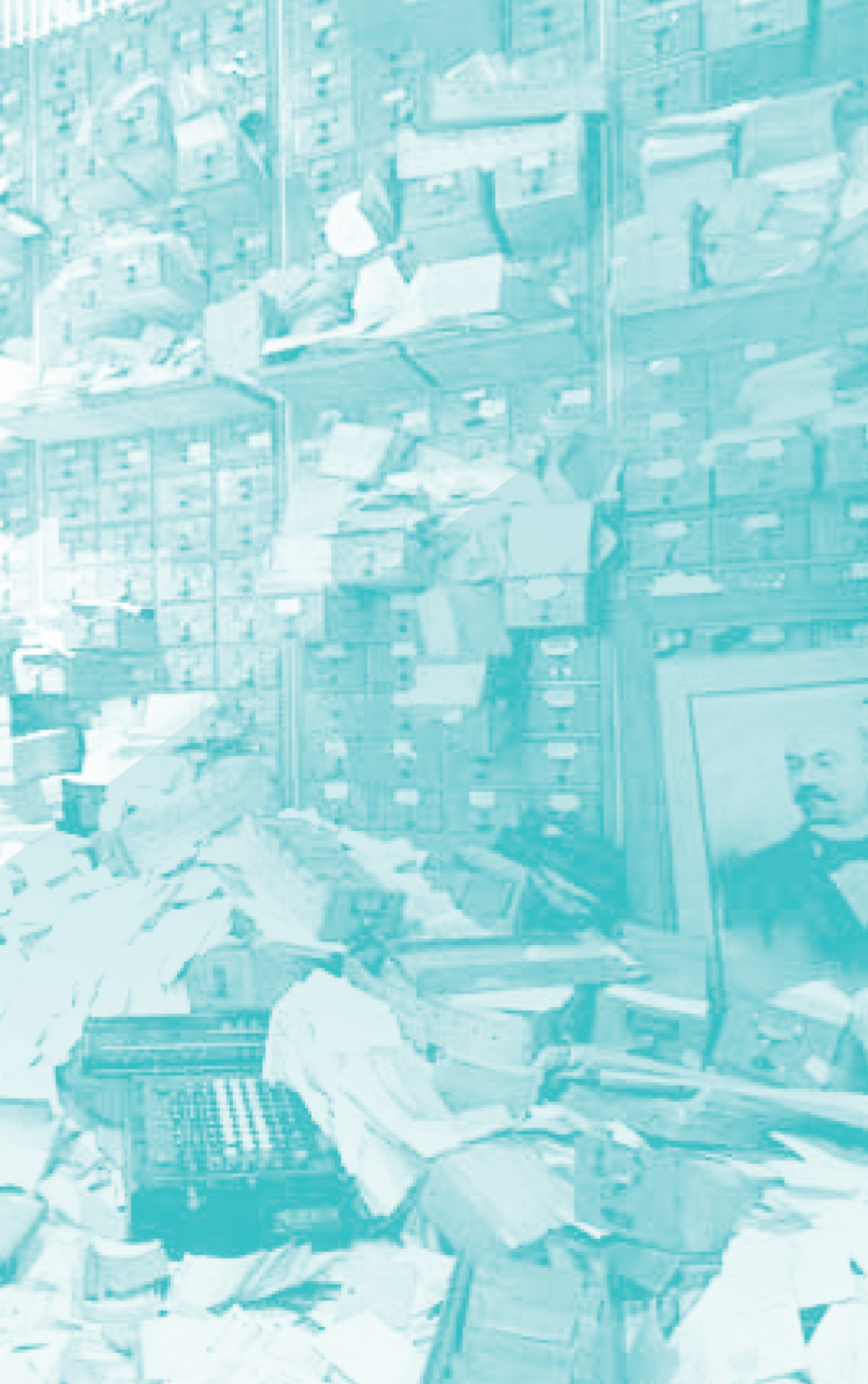

Book to the Future
a manifesto for book liberation

The Book Liberation Manifesto is an exploration of publishing outside of current corporate constraints and beyond the confines of book piracy. We believe that knowledge should be in free circulation to benefit humankind, which means an equitable and vibrant economy to support publishing, instead of the prevailing capitalist hand-me-down system of Sisyphean economic sustainability. Readers and books have been forced into pirate libraries, while sales channels have been monopolised by the big Internet giants which exact extortionate fees from publishers. We have three proposals. First, publications should be free-at-the-point-of-reading under a variety of open intellectual property regimes. Second, they should become fully digital — in order to facilitate ready reuse, distribution, algorithmic and computational use. Finally, Open Source software for publishing should be treated as public infrastructure, with sustained research and investment. The result of such robust infrastructures will mean lower costs for manufacturing and faster publishing lifecycles, so that publishers and publics will be more readily able to afford to invent new futures.

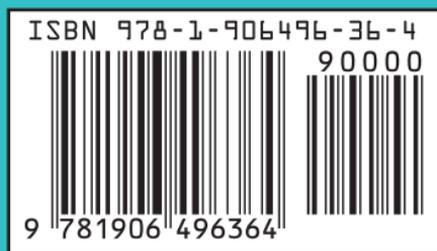

ISBN 978-1-906496-36-4